\documentclass[fleqn,usenatbib]{mnras}

\usepackage{mathptmx}

\usepackage[T1]{fontenc}
\usepackage{booktabs}
\usepackage{lipsum}
\usepackage{multicol}

\usepackage{graphicx}	
\usepackage{amsmath}	
\usepackage{amssymb}	

\title{SCSS-Net: Solar Corona Structures Segmentation by Deep Learning}

\author[Mackovjak et al.]{
\v{S}imon Mackovjak$^{1}$\thanks{E-mail: mackovjak@saske.sk},
Martin Harman$^{2}$,
Viera Maslej-Kre\v{s}\v{n}\'{a}kov\'{a}$^{2}$, and
Peter Butka$^{2}$
\\
$^{1}$Department of Space Physics, Institute of Experimental Physics, Slovak Academy of Sciences, Ko\v{s}ice, Slovakia\\
$^{2}$Department of Cybernetics and Artificial Intelligence, Faculty of Electrical Engineering and Informatics, Technical University of Ko\v{s}ice, Ko\v{s}ice, Slovakia\\
}

\date{Accepted 2021 September 03. Received 2021 August 11; in original form 2021 May 14}

\pubyear{2021}

\begin{document}
\label{firstpage}
\pagerange{\pageref{firstpage}--\pageref{lastpage}}
\maketitle

\begin{abstract}
Structures in the solar corona are the main drivers of space weather processes that might directly or indirectly affect the Earth. Thanks to the most recent space-based solar observatories, with capabilities to acquire high-resolution images continuously, the structures in the solar corona can be monitored over the years with a time resolution of minutes. For this purpose, we have developed a method for automatic segmentation of solar corona structures observed in EUV spectrum that is based on a deep learning approach utilizing Convolutional Neural Networks. The available input datasets have been examined together with our own dataset based on the manual annotation of the target structures. Indeed, the input dataset is the main limitation of the developed model's performance. Our \textit{SCSS-Net} model provides results for coronal holes and active regions that could be compared with other generally used methods for automatic segmentation. Even more, it provides a universal procedure to identify structures in the solar corona with the help of the transfer learning technique. The outputs of the model can be then used for further statistical studies of connections between solar activity and the influence of space weather on Earth.
\end{abstract}

\begin{keywords}
Sun: corona -- methods: data analysis -- techniques: image processing -- software: development
\end{keywords}

\section{Introduction}
\label{sec:introduction}

Solar activity has been quantified using various indices such as the sunspot number \citep{Clette2014}, the F10.7 index \citep{Tapping2013}, coronal index \citep{Rybansky2005}, and others \citep{Ermolli2014} over more than six decades. These indices represent integrated quantities for specific processes in the solar atmosphere as they are measured by ground-based instruments. Thanks to the most recent space missions such as e.g. SOHO \citep{Domingo1995}, Hinode \citep{Kosugi2007}, SDO \citep{Pesnell2012}, Solar Orbiter \citep{Muller2013}, and Parker Solar Probe \citep{Fox2016}, solar activity can be monitored in a much more detailed way. The temporal resolution in terms of seconds and spatial resolution down to hundreds of kilometers in multiple spectral bands enable individual events to be studied with particular consequences for Sun--Earth relations \citep{Muller2020, Lorincik2021}. However, the challenging task is to effectively process the huge amount of images that are automatically acquired with a very high resolution. This task requires sophisticated algorithms which are not as straightforward as the methods used for the creation of standard solar indices' time series.  

The initially used automated methods for detection of solar activity in the solar images are based on the predefined rules inferred from the appearance and usual characteristics of the structures in the solar atmosphere \citep{Henney2005,Krista2009,Perez2011}. However, it is not possible to capture all the unique solar structures using generally defined rules. Therefore, methods with more advanced mathematical algorithms were introduced. The Spatial Possibilistic Clustering Algorithm (SPoCA) \citep{Barra2009, Verbeeck2014} or Coronal Hole Identification via Multi-thermal Emission Recognition Algorithm (CHIMERA) \citep{Garton2018} have been found to be very effective and are widely used in online solar data visualization tools\footnote{\url{https://helioviewer.org}}$^,$\,\footnote{\url{https://www.solarmonitor.org}}. The SPoCA also provides entries for catalogues of coronal holes and active regions within the Heliophysics Events Knowledgebase (HEK) \citep{Hurlburt2012} that is commonly used in the SolarSoft \citep{Freeland1998} and SunPy \citep{sunpy_community2020} frameworks. As will be presented later, these algorithms still have limitations for the precise segmentation of structures in the solar corona. Due to the advances in computer vision in recent years, approaches based on machine learning techniques are able to extend the standard methods \citep{Aschwanden2010}. Conventional machine learning techniques as Support Vector Machine (SVM), Decision Tree, or Random Forest could improve the detection of coronal holes as they provide automated distinguishing from filaments in EUV solar images \citep{Reiss2015, Delouille2018}. On the assumption that the techniques based on Convolutional Neural Networks (CNN) are the most convenient techniques for image segmentation tasks \citep{lecun2015deep}, a new era in automated processing of solar images has begun. \cite{Illarionov2018} demonstrated that for the segmentation of coronal holes, CNN provides quantitatively comparable results as the SPoCA and CHIMERA algorithms. \cite{jarolim2021multi} provided a method called CHRONOS based on CNN with progressively growing network approach for robust segmentation of coronal holes.  

In what follows, we present our own \textit{SCSS-Net} model for the segmentation of solar corona structures based on a deep learning approach. The performance of the model is examined by segmentation of active regions and coronal holes while the applicability of the approach also for other corona structures is presented. The details about the data used, the annotations, and preprocessing of the data are presented in Section \ref{sec:data}. The deep learning model developed, its optimization, and metrics are described in Section \ref{sec:methods}. The results are discussed in Section \ref{sec:evaluation} and the conclusions are summarized in Section \ref{sec:conclusions}.

\section{Data}
\label{sec:data} 

The main requirement for the usage of deep learning techniques is a sufficient amount of the data needed for the learning process (see Section \ref{sec:methods}). The instrument Atmospheric Imaging Assembly (AIA) \citep{Lemen2012} onboard the Solar Dynamics Observatory (SDO) \citep{Pesnell2012} meets this requirement as it has produced $\sim$\,2\,TB of uncompressed data each day since spring 2010. With spatial resolution of 1.5\,arcsec and temporal resolution of 12\,sec it captures full-disk solar images of $4096\times4096$ pixels in 10 spectral channels. These unprecedented capabilities of the AIA have significantly advanced understanding of the variability of the solar upper atmosphere. There are ready-to-use SDO\,/\,AIA datasets for computer vision and machine learning applications with a total size of 284\,GB \citep{Kucuk2017} and 6.5\,TB \citep{Galvez2019}. Even more, the particular data can easily be accessed via the public Helioviewer API\footnote{\url{https://api.helioviewer.org}} \citep{Hughitt2012, Vorobyev2018} by specifying the date and AIA channel. In our work we used this last mentioned approach and data of AIA 171\,\AA\ and AIA 193\,\AA\ were downloaded (one image per channel per day). Only data from these two channels were used because our target structures were active regions and coronal holes that have the best visibility in spectral channels  171\,\AA\, and 193\,\AA, respectively. The images were downloaded in JP2 format and then converted to PNG format with a resolution of $1024\times1024$ pixels as this format and resolution are satisfactory for the intended purpose.

Before the deep learning process can start, the target structures need to be annotated (labeled). As will be discussed later, this is the most critical step of the whole process. If the input annotation of segmented structures is not precise on the pixel level, precise output segmentation prediction can not be expected. To prepare correct and precise annotations of target structures, it requires time consuming expertise for each solar image used. This expertise can be partially automated by using already available segmentation methods. However, after the application of these methods, the generated annotations need to be revised and only images with accurate annotation of all target structures can be used. In our work, we employed the SPoCA \citep{Verbeeck2014} algorithm to prepare an initial set of annotated coronal holes and active regions. For each solar image we obtained the pixel positions of the boundaries of the target structures. Then, the images were revised one by one utilizing the power of Zooniverse\footnote{\url{www.zooniverse.org}} crowd-sourcing platform. The trained students verified if all coronal holes and active regions had been correctly annotated by the color contours on solar corona images. Incorrect annotations were manually corrected or the images were excluded. For coronal holes, from the 900 reviewed images of AIA 193\,\AA\, only 27\,\% were identified as completely accurately annotated by the SPoCA algorithm. It was even fewer for active regions, where almost all annotated images needed some manual corrections. It is noted that the reviewed images were evenly selected from the 4-year period 2012--2016 to obtain high variability of target structures. By this procedure our own Custom dataset of annotated target structures was created. To assure the reproducibility of our deep learning approach, we also employed annotations of coronal holes and active regions generated by the SPoCA \citep{Verbeeck2014} and annotations of coronal holes generated by CHIMERA \citep{Garton2018} and Region Growth\footnote{\url{https://observethesun.com}} \citep{Tlatov2014} algorithms, without any additional expert revision. The SPoCA annotations were downloaded via HEK's web API by HEK module implemented in SunPy. Specifically, the HEKClient class \footnote{\url{https://docs.sunpy.org/en/stable/api/sunpy.net.hek.HEKClient.html\#sunpy.net.hek.HEKClient}} was employed with set arguments of time, type of event (i.e. 'CH' or 'AR') and source algorithm (i.e. 'SPoCA'). The API response contains the $X$ and $Y$ coordinates in pixel units of boundaries of the requested structure (attribute "hpc\_boundcc" in HEK Response object) and the time of event (attribute "event\_endtime" in HEK Response object). If the event time was slightly different than the AIA image time, the event boundaries were rotated by SunPy methods to map heliographic location at the observation time. The CHIMERA annotations were downloaded by our own script from the web repository\footnote{\url{https://solarmonitor.org/data/YYYY/MM/DD/meta/arm_ch_location_YYYYYMMDD.txt}, date example: 20181224} for all available dates. The obtained .txt text files contain the exact time of AIA observation and coordinates of coronal holes' boundaries in pixel units. To allow comparison of our results with \cite{Illarionov2018}, we also used exactly the same annotated dataset of coronal holes\footnote{\url{https://github.com/observethesun/coronal_holes/tree/mnras2018/data}} generated by Region Growth algorithm. The coordinates in pixel units of coronal holes and their boundaries are stored in .abp text files with date and time of AIA observations in the files' title. It is noted, AIA images were always linked with particular annotations according to the available exact date and time.

\begin{figure*}
    \centering
    \includegraphics[width=0.99\textwidth]{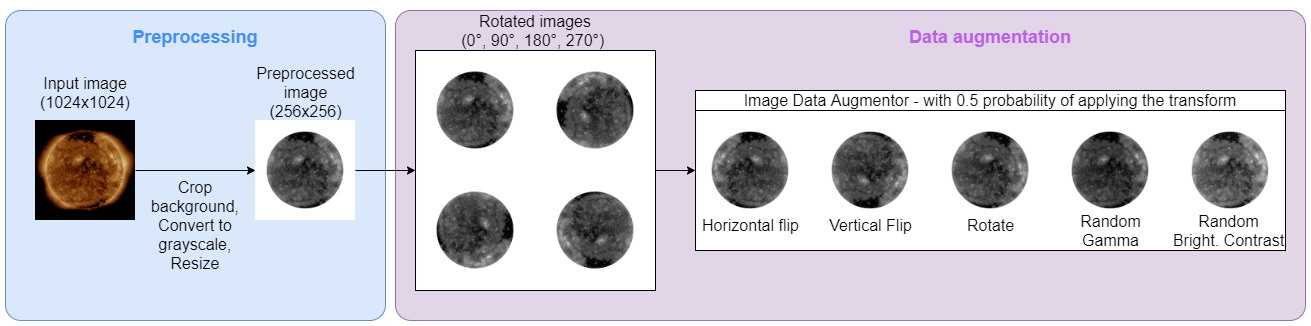}
    \caption{Example of data preparation process on the AIA 193\,\AA\, image acquired on June 30, 2018 at 03:46:16 UTC. Data preparation consists of preprocessing (blue box) and data augmentations (purple box). Using this process, for one input image we obtained four output images suitable for the following deep learning approach.}
    \label{fig:preprocessing}
\end{figure*}

The selected data were then preprocessed and augmented. The solar disk on AIA 193\,\AA\, images was cropped using the known center and diameter of the disk for each image. The black background was removed to prevent segmentation errors for coronal holes located on the edge of solar disk since the structures of coronal holes are characterized by dark pixels  \citep{Illarionov2018}. For AIA 171\,\AA\, images, this step is not required. Subsequently, the images were converted to grayscale and the resolution was reduced to $256\times256$ pixels. The preprocessing of AIA 193\,\AA\, images is summarized in Figure \ref{fig:preprocessing}, in the blue box. The preprocessing was applied to the entire dataset. To increase the number of images for the training process we applied data augmentation techniques. They are based on the fact that minor changes, such as flip, translation, or rotation, produce new images that might improve the overall success of CNN \citep{shorten2019survey, mikolajczyk2018data}. To create relevant new images, we used the affine transformation -- rotation. We rotated each image by 90, 180, and 270 degrees, thus obtaining a four times larger dataset. In addition, we used the Albumentations \citep{Buslaev2019} Python library for additional augmentations. Specifically, we used ImageDataAugmentor \citep{Tukiainen2019} that applies the following augmentations with a 50\,\% probability of applied transform: horizontal flip (flip the input horizontally around the y-axis), vertical flip (flip the input vertically around the x-axis), rotate (rotate the input by an angle from the range $(-45^{\circ}, 45^{\circ})$ selected randomly from the uniform distribution), random gamma (apply a random gamma distribution with a gamma limit range $(100,150)$), random brightness contrast (randomly change brightness and contrast of the input image in range of brightness $(0,0.2)$ and range of contrast $(0,0.4)$). Examples of data augmentation techniques used are presented in Figure~\ref{fig:preprocessing}, in the purple box. The output images of the preprocessing and augmentation process together with their labels (annotated positions of target coronal holes and active regions) were then used as described in Section \ref{sec:evaluation}.

\section{Deep learning approach}
\label{sec:methods} 

Image segmentation is the primary area of computer vision that specializes in selecting the class label for each pixel \citep{HARALICK1985100}. Most of the methods for image segmentation in the past applied techniques like thresholding \citep{sahoo1988survey}, clustering \citep{hartigan1979algorithm}, or edge detection \citep{canny1986computational}. In recent years, the most successful image segmentation techniques use deep learning approaches, specifically Convolutional Neuronal Networks (CNN) \citep{lecun2015deep}. CNN can be distinguished by the architecture used, which consists of hidden layers that connect the input and output layers of a deep neural network. The architecture used in our work is based on the Encoder-Decoder principle. For the encoding part of the CNN, pre-trained models such as AlexNet \citep{krizhevsky2017imagenet} or ResNet \citep{he2016deep}, are often used. The Encoder part provides the discriminative features of the original samples in compact, low-resolution space. The Decoder part then semantically projects the extracted features onto the pixel space (back in a higher resolution) to get the dense classification of pixels into selected classes. We have developed our own \textit{Solar Corona Structures Segmentation Network (\textit{SCSS-Net})} that is inspired by U-Net \citep{ronneberger2015u} and is based on Encoder-Decoder architecture of CNN.

\subsection{Architecture of SCSS-Net}
\label{sec:archiitecture} 
The encoder of the \textit{SCSS-Net} architecture consists of five convolutional blocks. The input to \textit{SCSS-Net} is a solar corona image (output from the preparation process displayed in Figure \ref{fig:preprocessing}) of size $256\times256$ pixels. We choose this resolution as optimal because the calculations with this image resolution is not demanding in terms of specialized hardware and can be performed on a standard computer (we used CPU Intel i5 8300H, RAM 32GB) with GPU (we used NVIDIA GeForce GTX 1060 Max-Q 6GB). Each convolutional block consists of two convolutional layers with a kernel window size of $3\times3$ pixels. The padding is set evenly so the output has the same dimensions as the input. The number of filters in the first convolutional block starts at 32, and the number of filters gradually doubles in each subsequent block. The computation of convolutional layers outputs is given by:
    \begin{equation}
     S(i,j)=(K*I)(i,j)=\sum_m \sum_n I(m,n)\;K(i-m,j-n),
    \end{equation}
where $m$ is the number of rows and $n$ is the number of columns for $I$ as the input image or output of a particular layer, $K$ is a kernel (filter), and $S(i,j)$ represents the result of convolution for $i$-th row and $j$-th column \citep{Goodfellow-et-al-2016}. The convolutional layer is followed by a batch normalization layer, which speeds up and stabilizes convergence of the training process \citep{normalization}. Batch normalization applies a transformation that maintains the mean output close to 0 and the output standard deviation close to 1. In all blocks (convolutional and deconvolutional), we used a rectified linear function (ReLU) as an activation function defined as \citep{nair2010rectified}:
    \begin{equation}\label{e:relu}
    f(x) =  \max (0, x),
    \end{equation}
where $x$ is an input value. The max-pooling layers are present between the convolutional blocks. They provide a reduction step that is important for the feature extraction. Max-pooling is a pooling operation that calculates the maximum value in each patch of each feature map. In our case, we used a pool size of $2\times2$ pixels. For the last three convolutional blocks, we used regularization in the form of a dropout layer. Dropout \citep{srivastava2014dropout} randomly sets the input units to 0 at each step (training example) during training time, excluding them from training and preventing excessive overfitting. Dropout has a hyperparameter that specifies the probability at which the layer units are omitted.

\begin{figure*}
    \centering
    \includegraphics[width=0.9\textwidth]{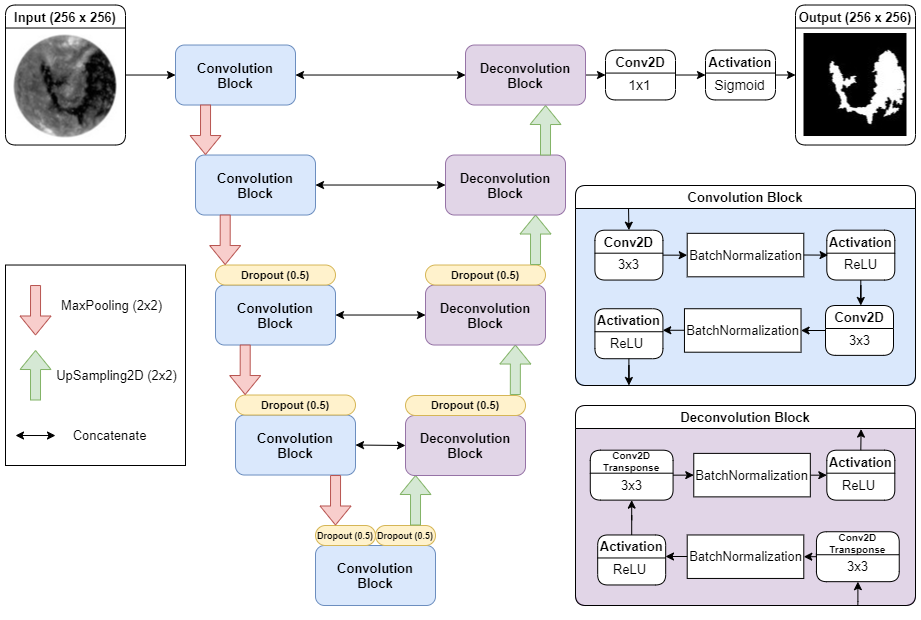}
    \caption{\textit{SCSS-Net} model architecture inspired by U-Net \citep{ronneberger2015u} configuration. The input to the \textit{SCSS-Net} is a solar corona image of size $256\times256$ pixels. \textit{SCSS-Net} is based on the Encoder-Decoder principle. The encoder, the left part of the figure, is composed of five convolution blocks. The red arrows indicate max-pooling layers with a pool size of $2\times2$ pixels. Convolutional blocks contain two convolutional layers, batch normalization and ReLU activation function. The convolutional layers in blocks double their number of filters for each depth level - from $32$ at the start to $512$ at the bottom of the encoder part. The decoder, the right part of the figure, contains four deconvolutional blocks. The green arrows indicate up-sampling layers with a pool size of $2\times2$ pixels. Deconvolution blocks consist of two transpose convolution layers, batch normalization, and the ReLU activation function. Deconvolutional layers in blocks contain a number of filters progressively halving from $256$ to $32$. The deeper layers of the network with more filters also contain dropout regularization layers, both in the encoder and decoder parts. Black arrows indicate concatenation layers, which concatenate feature maps within the same level. This operation helps to give the localization information from encoder to decoder. The output layer consists of the convolutional layer with the sigmoid activation function. The output of \textit{SCSS-Net} is a segmentation mask with size of $256\times256$ pixels.}
    \label{fig:scss_net}
\end{figure*}

The decoder part in \textit{SCSS-Net} architecture consists of four deconvolution blocks. Each block contains two transposed convolution layers with a kernel window size of $3\times3$ pixels, batch normalization, and ReLU activation function. The number of filters in the deconvolutional block starts at 256, and for each subsequent block the number of filters is reduced by half. These blocks are connected by an UpSampling layer that repeats the rows and columns of data. We also added a dropout regularization layer behind the first two deconvolutional blocks. Each UpSampling layer is followed by a concatenation of feature maps with the same level from encoder part. This helps to improve localization information between encoder and decoder. The output layer consists of the convolutional layer with a kernel size of $1\times1$ pixel and the sigmoid activation function:
    \begin{equation}\label{e:sigmoid}
        f(x) = \frac{1}{1+e^{-x}},
    \end{equation}
which classifies each pixel of the image as 0 or 1 based on set threshold. The result for the whole network is the prediction of a class for each image pixel. This can be visualized as the output segmentation mask \citep{nwankpa2018activation}. The detailed architecture of the \textit{SCSS-Net} is summarized in Figure \ref{fig:scss_net}.

\subsection{Optimizations and evaluation metrics}
The optimization methods improve the attributes of the neural network, such as weights and the learning rate.  The main goal is to reduce the error (loss) of the network prediction. We tested several optimizers during the hyperparameter fine-tuning phase: Root Mean Square Propogation (RMSProp, \cite{hinton2012neural}), Stochastic gradient descent (SGD, \cite{Robbins2007ASA}), Adaptive subgradient methods (Adagrad, \cite{duchi2011adaptive}) and Adaptive moment estimation (Adam, \cite{kingma2014adam}). Our fine-tuning experiments showed that the Adam optimizer provides best results for our model. Adam computes the learning estimation for each parameter. Adam also stores an exponentially decaying average of past squared gradients $v(t)$ and an exponentially decaying average of past gradients $m(t)$. Both moving averages are initialized to 0, which leads to the moments' estimation biased towards zero. Such a situation occurs mostly during the initial phases when decay parameters ($\beta$1, $\beta$2) have values close to 1. Such~bias can be removed using modified estimations $\hat{m_{t}}$ and $\hat{v_{t}}$:
\begin{equation}\label{e:adam}
    \hat{m_{t}} = \frac{m_{t}}{1 - \beta^{t}_{1}}, \qquad \hat{v_{t}} = \frac{v_{t}}{1 - \beta^{t}_{2}}.
\end{equation}
The main goal of the Adam optimization algorithm is to optimize stochastic objective function $f(\theta)$, where parameter $\theta$ represents weights and biases. The parameters $\theta$ are updated according to the formula:
\begin{equation}
    \theta_{t+1}   = \theta_t  \frac{\alpha}{\sqrt{\hat{v}_t} +\epsilon} \cdot \hat{m}_t.
\end{equation}
The default settings for the parameters are $\beta_1=0.9$ and  $\beta_2=0.999$, learning rate $\alpha = 0.001$ and parameter $\epsilon=10^{-8}$ \citep{kingma2014adam}.

The neural network updates weights between neurons using the backpropagation of error algorithm, which was first introduced in \cite{rumelhart1986learning}. The function of error that is needed to be minimized is called the cost function also known as loss function or error function \citep{Goodfellow-et-al-2016}. In our task with a binary classification of a pixel belonging or not belonging to the target class, we used {Binary Cross-Entropy} (BCE) as the loss function. It can be defined as:
\begin{equation}\label{e:bce} 
    \mathrm{BCE} = -\left ( y\log \left ( \hat{y} \right ) +
    \left ( 1 - y \right )\log \left ( 1 - \hat{y}
    \right ) \right ), 
\end{equation} 
where $y$ represents the ground truth (target label) and $\hat{y}$ represents the predicted value \citep{murphy2012machine}.

To evaluate the performance of our developed \textit{SCSS-Net}, we used the Jaccard index \citep{legendre2012numerical} and Dice coefficient \citep{sorensen1948method} metrics. The Jaccard Index is called Intersection over Union (IoU) for image segmentation tasks \citep{Goodfellow-et-al-2016}. In more detail, IoU is the overlap between the predicted segmentation and the ground truth divided by the area of union between the predicted segmentation and the ground truth.  The Dice coefficient, also known as F1 score in data analysis \citep{sorensen1948method}, is a statistical tool that measures the similarity between two sets of data. Both IoU and Dice can be defined as follows:
\begin{equation}\label{e:iou}
    \mathrm{IoU} = \frac{\mathrm{TP}}{\mathrm{TP + FP + FN}},
\end{equation}

\begin{equation}\label{e:dice}
    \mathrm{Dice} = \frac{\mathrm{2TP}}{\mathrm{2TP + FP + FN}},
\end{equation}
where: TP (true positive) – indicates the amount of correctly segmented pixels, i.e., the number of pixels for which the prediction class and ground truth class is the same, FP (false positive) - indicates the amount of predicted object mask pixels with no association to ground truth object mask pixels, FN (false negative) – indicates the amount of ground truth object mask pixels with no association to predicted object mask pixels. It is important to note, that if some pixels are not labeled as ground truth but they should be (which is a very common scenario) the FP value can be high and therefore the overall metrics can not reach the maximal values in their $(0,1)$ interval.

\subsection{Regularization and fine-tunning} 
\label{subsec:finetune}
The next step in \textit{SCSS-Net} development was the use of regularization techniques to prevent overfitting. In our experiments, we used Dropout \citep{srivastava2014dropout} as the primary regularization technique. Also, we used a checkpoint\footnote{\url{https://keras.io/api/callbacks/model\_checkpoint}} to save the model that achieved a minimum loss on the validation set during training, which helps to select the best model without overfitting issues.

Fine-tuning of the model hyperparameters is a crucial step to improve the results \citep{li2020rethinking}. We used a grid search approach and varied the image's input size, dropout rate, batch size, and tested various optimizers during the training. The fine-tuning values used on \textit{SCSS-Net} training on the Custom dataset for segmentation of coronal holes are listed in Table \ref{tab:finetune}. Our model's best setting of hyperparameters was as follows: image input size $256\times256$ pixels, dropout rate $0.5$, batch size $20$, and optimizer Adam.

\begin{table}
	\centering
	\caption{Hyperparameters used during fine-tuning of the \textit{SCSS-Net} model.}
	\label{tab:finetune}
	\begin{tabular}{lc} 
		\toprule
		 \textbf{Hyper-parameters} & \textbf{Values}\\
		\midrule
		 Image size & \texttt{[128 $\times$ 128, 256 $\times$ 256, 512 $\times$ 512]} \\
		 Dropout rate &  \texttt{[0.5, 0.25]} \\
		 Batch Size &  \texttt{[20, 32]}\\
		 Optimizer & \texttt{['SGD', 'RMSProp', 'Adam', 'AdaGrad']}\\
		\bottomrule
	\end{tabular}
\end{table}

\begin{figure}
    \centering
    \includegraphics[width=0.48\textwidth]{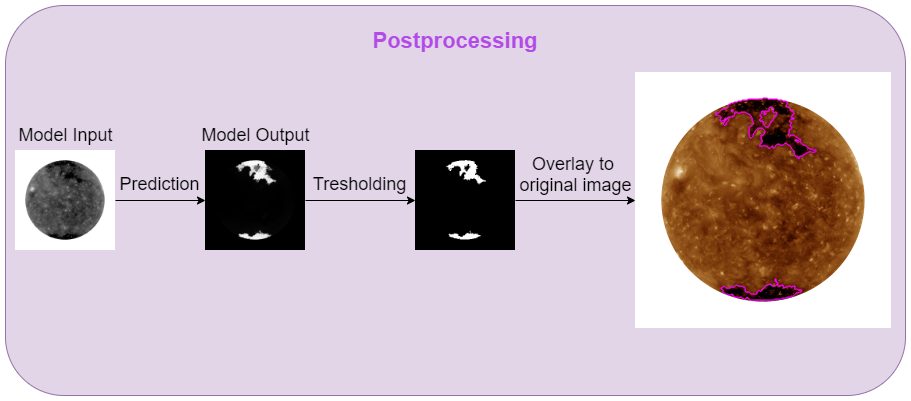}
    \caption{Example of the post-processing procedure. The outputs of the \textit{SCSS-Net} model are the probabilities of each pixel belonging to the target class. These probability values are then compared with the threshold value to get final segmentation mask. This mask can be then applied as an overlay on the input image.}
    \label{fig:postprocessing}
\end{figure}

The prediction from \textit{SCSS-Net} provides the class probability for each pixel. We adjusted the prediction based on the threshold, which we set to $0.5$. Thresholding divides an image into a foreground and background. A specified threshold value separates pixels into one of two levels to isolate predictions. This step provides a binary output segmentation mask, where $1$ represents the pixels of the predicted solar corona structure and $0$ represents the pixels outside of the predicted structure. The obtained masks can be then overlaid on the input images to visualize precision of the prediction. The whole post-processing process is displayed in Figure \ref{fig:postprocessing}.

It is important to mention that our deep learning model \textit{SCSS-Net} was developed in the Python programming language \citep{van1995python} using the Tensorflow \citep{abadi2016tensorflow} and Keras \citep{chollet2015keras, gulli2017deep} libraries. The source code, together with a detailed description is available in the Jupyter notebook that forms the online\footnote{\url{https://github.com/space-lab-sk/scss-net}} material to this article.

\section{Results and Discussion}
\label{sec:evaluation}

\subsection{Segmentation of coronal holes} 
\label{subsec:ch}
We trained our own \textit{SCSS-Net} based on a deep learning approach (Section \ref{sec:methods}) using datasets with four types of coronal holes' annotations -- Custom, SPoCA, CHIMERA, and Region Growth (Section \ref{sec:data}). The models trained for each annotation source were evaluated on validation datasets that represent $10\%$ of randomly selected data from particular initial dataset. The performance of the models during the training was also monitored using these validation sets. Table \ref{tab:size} summarizes the number of images used with annotation in initial datasets, augmented datasets (with applied rotation), train datasets, and validation datasets. For example, the initial dataset with Custom annotations contained $537$ AIA 193\,\AA\, images. By augmentation -- rotation of images by $90$, $180$, and $270$ degrees we obtained $2148$ images. Then the data were split in the ratio $90\%$ (i.e. $1933$ images) to $10\%$ (i.e. $215$ images) for the training set and validation set, respectively. The data in training set were then modified using ImageDataAugmentor random transformations as described in Section \ref{sec:data}.

\begin{table}
	\centering
	\caption{Sizes of datasets for four types of annotations: Custom, SPoCA, CHIMERA, and Region Growth. The first column represents the size of the initial dataset. The second column contains four times more images than the initial dataset - after rotating by 90, 180, and 270. The third column is the number of images in the training set to which ImageDataAugmentor is applied. The validation set is 10\% of the dataset after rotation.}
	\label{tab:size}
	\begin{tabular}{lcccc} 
		\toprule
		 & \textbf{Initial}  & \textbf{Rotated} & \textbf{Train set} & \textbf{Validation set} \\
		\midrule
		Custom   &  537  & 2148 & 1933 & 215 \\
		SPoCA   &   537  & 2148 & 1933 & 215\\
		CHIMERA  &  653  & 2612 &  2351 & 261\\
		Region Growth  &  1968 & 7872  &  7085 & 787 \\
		\bottomrule
	\end{tabular}
\end{table}

\begin{figure}
    \centering
    \includegraphics[width=0.36\textwidth]{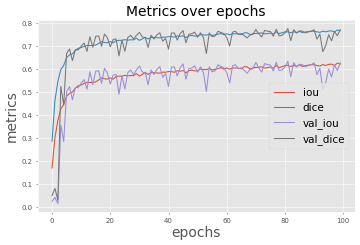}
    \includegraphics[width=0.36\textwidth]{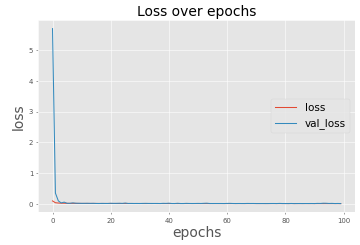}
    \caption{Training process of \textit{SCSS-Net} for segmentation of coronal holes using Custom annotations. \textit{Top:} The evolution of metrics IoU and Dice on training and validation set over 100 epochs. \textit{Bottom:} The evolution of the loss function for training and validation sets over 100 epochs.}
    \label{fig:training_ch}
\end{figure}

The learning process for segmentation of coronal holes using the Custom dataset is presented in Figure \ref{fig:training_ch}. The values of metrics IoU and Dice for training and validation sets are displayed together with the values of loss function during 100\,epochs. The best model was saved after the epoch that achieved minimum loss on the validation set. Then the best model was used to provide predictions for each image in the validation set and the outcome metrics were averaged over the whole set. The results for each validation dataset is presented in Table \ref{tab:ch1}. To provide insight to these results, we created a sequence of 4 images that consists of the input image for \textit{SCSS-Net}, ground truth (i.e. annotation mask), output of \textit{SCSS-Net} (i.e. the predicted segmentation mask), and the predicted segmentation mask as an overlay on the input image. Examples of two such image sequences are displayed in Figure \ref{fig:val_custom} while the \textit{SCSS-Net} was trained on the dataset with Custom annotations. The metrics IoU and Dice for particular image sequences are also displayed. The example of the \textit{SCSS-Net} performance trained on the dataset with SPoCA annotations is presented in Figure \ref{fig:val_spoca}. In this case, it is visible that the annotations seem to be undersized and the whole area of the coronal holes is not correctly segmented. Conversely, for the dataset with CHIMERA annotations in Figure \ref{fig:val_chimera}, the annotations seem to be oversized. The best results are achieved for the model trained on the dataset with Region Growth annotations (Figure \ref{fig:val_region}). The lowest metrics' values are for the model trained on the dataset with SPoCA annotations. This might be explained by imperfections in the ground truth annotations generated by SPoCA. The predictions by our model seem to correctly segment a bigger area of coronal holes than the SPoCA annotations and this increases the number of false positive pixels and consequently decreases the IoU and Dice values (see Eq. \ref{e:iou} and Eq. \ref{e:dice}). On the other hand, annotations from other sources seem to be more consistent therefore also the predictions have higher performance.

\begin{table}
	\centering
	\caption{\textit{SCSS-Net} metrics for segmentation of coronal holes for four types of annotations used for training. The results represent the performance of the best model on the validation set. The values of Dice and IoU are averages over the whole validation set.}
	\label{tab:ch1}
	\begin{tabular}{lccccr} 
		\toprule
		\textbf{Train dataset}  & \textbf{Validation set}  & \textbf{Dice  } & \textbf{IoU} \\
		\midrule
		Custom  & 215  & 0.85 &  0.74  \\
		SPoCA   & 215  & 0.78 & 0.64 \\
		CHIMERA & 261  & 0.86 &  0.76  \\
		Region Growth   &  787   & 0.88 & 0.78  \\
		\bottomrule
	\end{tabular}
\end{table}

\begin{figure}
    \centering
    \includegraphics[width=0.48\textwidth]{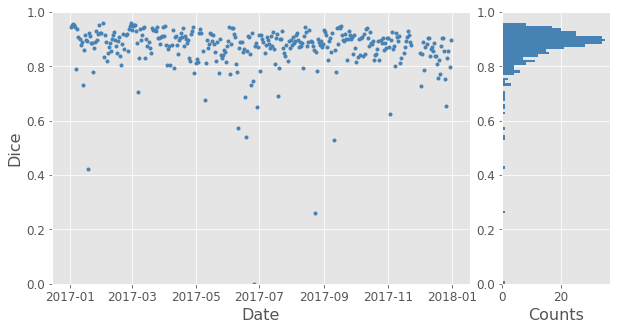}
    \caption{The values of Dice score for the model trained on Region Growth dataset for each image in the Test set during the year 2017 (\textit{left}). Distribution of Dice score values in the Test set (\textit{right}). These panels extend the information of the averaged Dice score presented in Table \ref{tab:ch2b}, last row.}
    \label{fig:ch_dice}
\end{figure}

\begin{figure}
    \centering
    \includegraphics[width=0.48\textwidth]{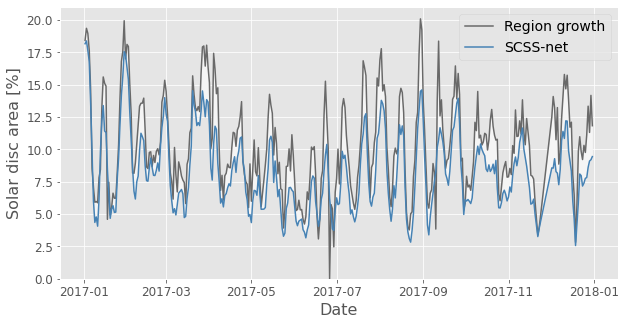}
    \caption{The area of segmented coronal holes in \% of total solar disc area by Region Growth reference annotations and by \textit{SCSS-Net} model. The visualization is complementary to Figure \ref{fig:ch_dice} and represents the robustness of the developed SCSS-net.}
    \label{fig:ch_area}
\end{figure}

To be able to compare the results of these models objectively, we tested the pre-trained models on completely unseen datasets from the year 2017. Data from this year were not used in any of previous training or validation sets. As a reference we used SPoCA and Region Growth annotations to generate ground truth annotation masks. The predictions generated by four pre-trained models discussed above were compared against these two reference annotations. For models trained on a one type of annotation and tested on another type, we talk about transfer learning - pre-trained model approach \citep{transferLerning, Goodfellow-et-al-2016}. We trained the model (i.e. set weights and biases for each neuron) using one dataset and then used this pre-trained model on completely unseen data. The results of these tests are listed in Table \ref{tab:ch2} and in Table \ref{tab:ch2b} for SPoCA and Region Growth reference annotations, respectively. To provide insight to these results, the examples of image sequences with quantified IoU and Dice values are presented in Figures \ref{fig:test1_custom}-\ref{fig:test1_region} for SPoCA reference annotations and in Figures \ref{fig:test2_custom}-\ref{fig:test2_region} for Region Growth reference annotations. The corresponding image sequence pairs (in Figures \ref{fig:test1_custom} \& \ref{fig:test2_custom}, \ref{fig:test1_spoca} \& \ref{fig:test2_spoca}, \ref{fig:test1_chimera} \& \ref{fig:test2_chimera}, \ref{fig:test1_region} \& \ref{fig:test2_region}) contain the same input image,  predicted segmentation mask, and overlaid image. It is due to the fact that particular pre-trained models were the same (e.g. the same Custom pre-trained model for coronal holes was used to generate predictions in Figures \ref{fig:test1_custom} \& \ref{fig:test2_custom}). The image sequence pairs differ in the reference annotation masks and therefore also in values of Dice and IoU metrics (e.g. in Figure \ref{fig:test1_custom} \textit{(first row)} Dice: 0.77, IoU: 0.62 and in Figure \ref{fig:test2_custom} \textit{(first row)} Dice: 0.91, IoU: 0.84). The metrics for Custom, CHIMERA, and Region Growth pre-trained models are much higher for Region Growth reference annotations (Table \ref{tab:ch2b}) than for SPoCA reference annotations (Table \ref{tab:ch2}). It is because the Region Growth algorithm is more precise for annotation of coronal holes than the SPoCA algorithm. Therefore the number of false positive pixels in prediction mask is lower for Region Growth than for SPoCA reference annotations. The results presented in Table \ref{tab:ch2b} and in Figures \ref{fig:test2_custom}-\ref{fig:test2_region} demonstrate very high performance of our \textit{SCSS-Net} also on the test dataset. The averaged value of Dice score for Region Growth annotations is by $\sim$\,7\,\% higher than the result obtained by other authors \citep{Illarionov2018}. To visualize the performance of \textit{SCSS-Net} to segment coronal holes on images in test set, we also created a movie from 353 AIA 193\,\AA\ images from year 2017. This model was trained on Region Growth dataset. The movie is available within online\footnote{\url{https://github.com/space-lab-sk/scss-net}} material to this article. The Dice score for each image used in movie is displayed in Figure \ref{fig:ch_dice}. The value of Dice score is higher than 0.8 for 90\% of used images. This demonstrate high stability of \textit{SCSS-Net} performance on completely previously unseen data. The robustness of \textit{SCSS-Net} is also confirmed in Figure \ref{fig:ch_area}. The area of segmented coronal holes is compared for reference annotations and \textit{SCSS-Net} predictions. It is interesting to note, if we consider the Region Growth annotations as the ground truth and the SPoCA annotations as the predictions and then calculate the metrics for the same test dataset as it is in Table \ref{tab:ch2b}, we obtain Dice = 0.60 and IoU = 0.43.

\begin{table}
	\centering
	\caption{Performance of \textit{SCSS-Net} pre-trained models for coronal holes segmentation against the SPoCA reference annotations. The test set consists of input images from the year 2017, only.}
	\label{tab:ch2}
	\begin{tabular}{lccccr}
		\toprule
		\textbf{Train dataset} & \textbf{Test set} & \textbf{Dice  } & \textbf{IoU} \\
		\midrule
		Custom  & 353 & 0.58 & 0.41  \\
		SPoCA  & 353 & 0.72 & 0.57 \\
		CHIMERA & 353 & 0.61 & 0.44  \\
		Region Growth  & 353  & 0.64 & 0.47 \\
		\bottomrule
	\end{tabular}
\end{table}

\begin{table}
	\centering
	\caption{Performance of \textit{SCSS-Net} pre-trained models for coronal holes segmentation against the Region Growth reference annotations. The test set consists of input images from the year 2017, only.}
	\label{tab:ch2b}
	\begin{tabular}{lccccr}
		\toprule
		\textbf{Train dataset} & \textbf{Test set} & \textbf{Dice} & \textbf{IoU} \\
		\midrule
		Custom  & 353 & 0.83 & 0.71  \\
		SPoCA  & 353 & 0.43  & 0.28 \\
		CHIMERA & 353 & 0.85  & 0.73   \\
		Region Growth  & 353  & 0.88 & 0.78 \\
		\bottomrule
	\end{tabular}
\end{table}

\begin{figure*}
    \centering
    \includegraphics[width=0.99\textwidth]{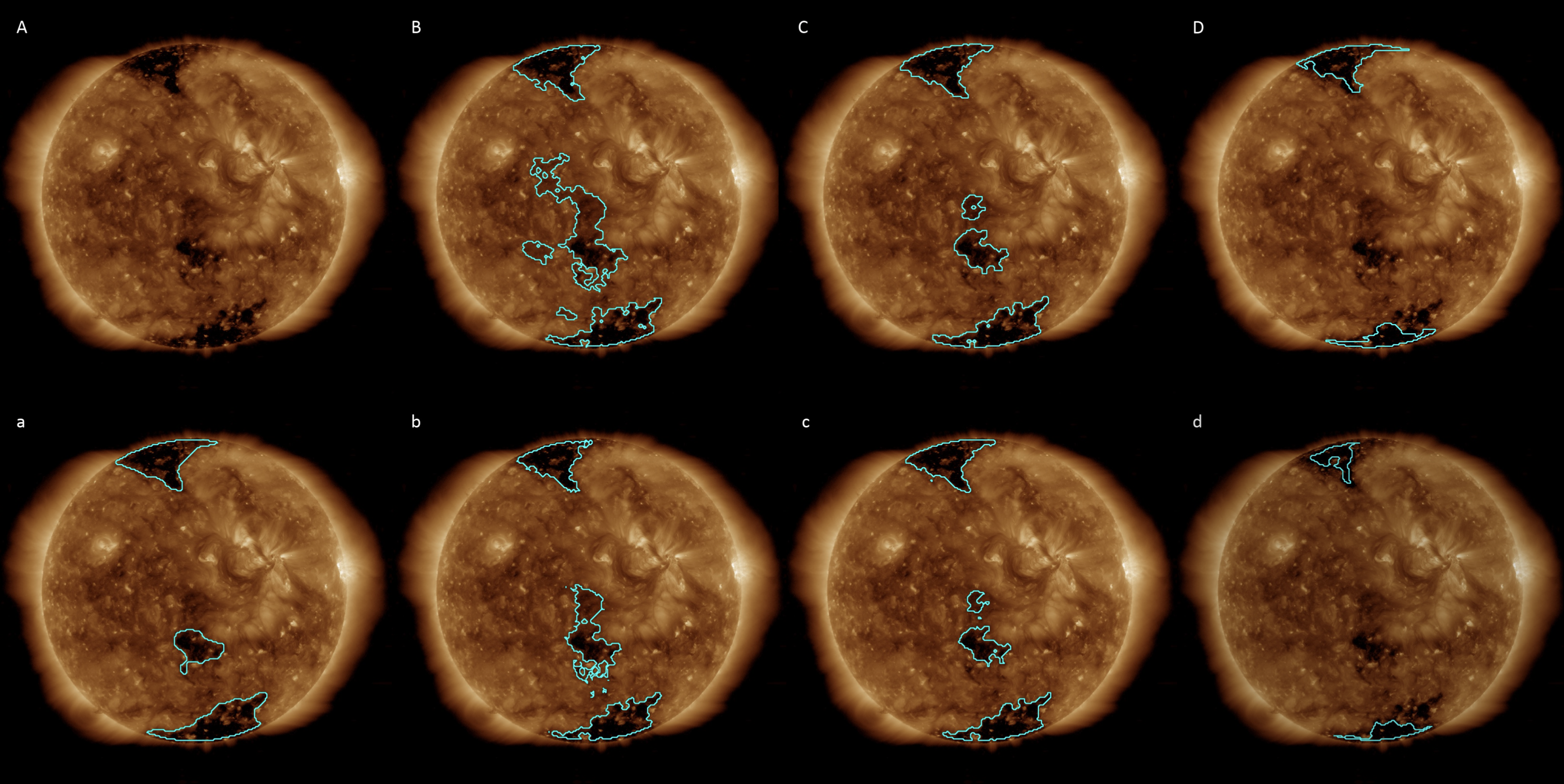}
    \caption{Segmentation of the AIA 193\,\AA\ image from 28 June 2018 at 2:42:28 using various methods. First row: Segmentation using commonly used algorithms: (A) there is no Custom annotation for this date, (B) contours using CHIMERA annotations, (C) contours by Region Growth annotation, (D) contours using SPoCA annotation. Second row: Segmentation using \textit{SCSS-Net} trained on four datasets: (a) Custom dataset, (b) CHIMERA dataset, (c) Region Growth dataset, (d) SPoCA dataset.}
    \label{fig:performence}
\end{figure*}

The developed \textit{SCSS-Net} model provides reasonable results in direct comparison with previously mentioned and commonly used algorithms for coronal holes segmentation. In Figure \ref{fig:performence}, we have compared segmentation of AIA 193\,\AA\ image acquired on 28 June 2018, at 02:42:28 obtained by commonly used algorithms (first row) with the \textit{SCSS-Net} segmentation (second row). The results of this experiment confirm that the input dataset is the main limitation of the developed deep learning model's performance. If the ground truth annotations contain a lot of imperfections, the deep learning model outputs contain them, too. However, if the input imperfections are not significant, they might be reduced during the deep learning process and might be not present in the output predictions. This means that the deep learning approach can learn to extract features better than the original algorithm. For example, when learning on a SPoCA dataset, it marks only the darkest (smaller) areas but doesn't indicate a place that isn't a coronal hole. Thus, the model, in this case, minimizes false negative issues (see Figure \ref{fig:performence} images (D)-(d)). On the other hand, the CHIMERA dataset is the opposite case. In the original dataset areas that are not coronal holes, are often marked as coronal holes. \textit{SCSS-Net} can suppress this error, thus minimizing false positives, as presented in Figure \ref{fig:performence} -- image (B)-(b). 

\subsection{Segmentation of active regions} 
\label{subsec:ar}

\begin{figure}
    \centering
    \includegraphics[width=0.36\textwidth]{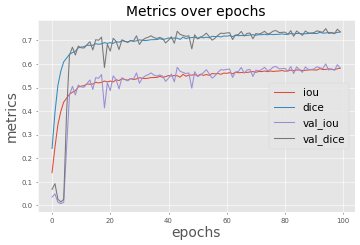}
    \includegraphics[width=0.36\textwidth]{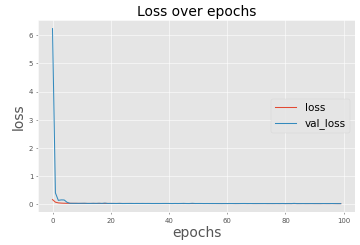}
    \caption{Training process \textit{SCSS-Net} for segmentation active areas on the Custom dataset. Right: The evolution of metrics (IoU, Dice) on train and validation set over epochs. Left: The evolution of loss function over epochs.}
    \label{fig:training_ar}
\end{figure}

The developed \textit{SCSS-Net} architecture, without any change of hyperparameters, can be used for segmentation of other solar corona structures. This is the advantage of a transfer learning - develop model approach. The only step that is needed is to re-train the model with the new input images and ground truth annotations. This approach was investigated using segmentation of active regions on the solar corona images. We utilised three datasets with the following annotations: Custom, SPoCA, and the combination of both. The re-trained models were tested on a validation set -- $10\%$ of the augmented dataset (with applied rotation), and also on test set with SPoCA reference annotations from the year 2016. The test set was not used in training process at all. Table \ref{tab:size_ar} summarizes the number of images in the initial dataset, augmented dataset (with applied rotation), training set, and validation set. The learning process for segmentation of active regions using the Custom dataset (IoU, Dice and Loss) during $100$ epochs is displayed in Figure \ref{fig:training_ar}. The best model was saved after the epoch that achieved a minimum loss on the validation set. 

\begin{table}
	\centering
	\caption{The first column represents the size of the initial dataset. The second column contains four times more images than the initial dataset - after rotating by 90, 180, and 270. The third column is the number of images in the training set to which ImageDataAugmentor is applied. The validation set is 10 percent of the dataset after rotation.}
	\label{tab:size_ar}
	\begin{tabular}{lcccc} 
		\toprule
		 & \textbf{Initial}  & \textbf{Rotate} & \textbf{Train set} & \textbf{Validation set} \\
		\midrule
		Custom  & 533  & 2132 & 1919 &  213 \\
		SPoCA   &  1091   &  4364 & 3928 & 436 \\
		SPoCA + Custom  &  1607  & 6428  & 5785  & 643\\
		\bottomrule
	\end{tabular}
\end{table}

The results of the best models for each input dataset are listed in Table \ref{tab:ar1}. The best averaged values of Dice and IoU metrics over the whole validation set were achieved by \textit{SCSS-Net} trained on a Custom dataset. The combination of the SPoCA and Custom datasets achieved numerically the worst model performance. However, the numerical results might be misleading because SPoCA dataset often contains false positive cases that the model has been able to suppress. Thus, once again we would like to highlight the image sequences attached in the appendix in Figures \ref{fig:ar_val_custom}-\ref{fig:ar_val_comb}.

\begin{table}
	\centering
	\caption{\textit{SCSS-Net} metrics for segmentation of active regions for three types of annotations used for training. The results represent the performance of the best model on the validation set. The values of Dice and IoU are averages over the whole validation set.}
	\label{tab:ar1}
	\begin{tabular}{lccccr} 
		\toprule
		\textbf{Train dataset} & \textbf{Validation set} & \textbf{Dice  } & \textbf{IoU} \\
		\midrule
		Custom  & 213 & 0.81 & 0.68  \\
		SPoCA & 436 & 0.73 &  0.57  \\
		SPoCA + Custom  & 643 & 0.52 & 0.35 \\
		\bottomrule
	\end{tabular}
\end{table}

\begin{figure*}
    \centering
    \includegraphics[width=0.99\textwidth]{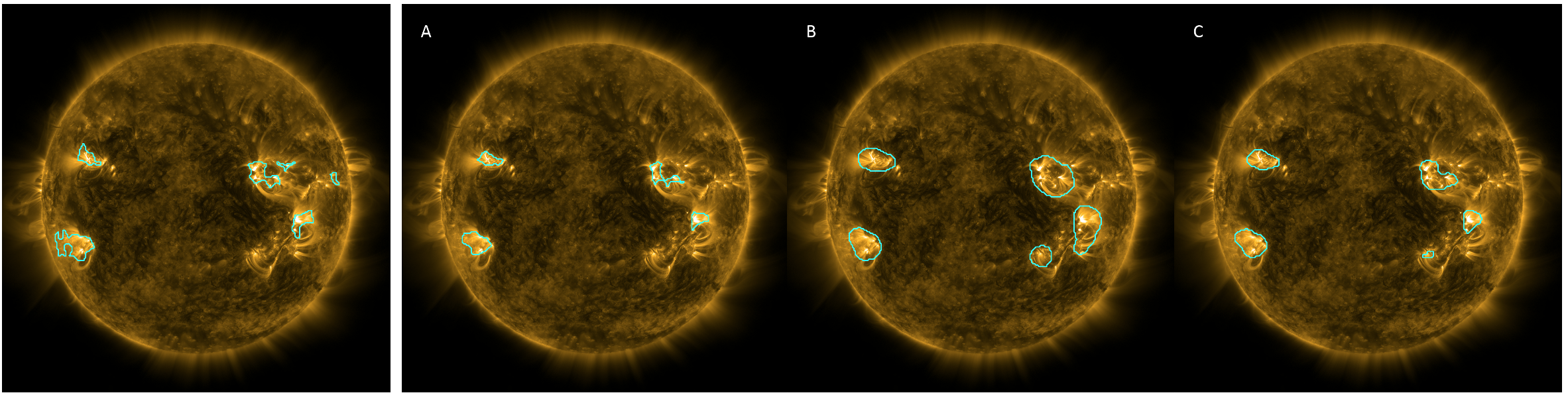}
    \caption{The image of solar corona from AIA 171\,\AA\, on 26 January 2016, 23:00:10:34. The first image (on the left) is overlaid by the contours generated using the SPoCA algorithm. The following  images are overlaid using three predictions of \textit{SCSS-Net} trained on: (A) -- SPoCA dataset, (B) -- Custom dataset, (C) -- the combination of SPoCA and the Custom dataset.}
    \label{fig:ar_predictions}
\end{figure*}

In Figure \ref{fig:ar_predictions}, there are four images of the Sun with contours associated to the position of active regions. The contours on the first image were obtained using the SPoCA algorithm, while the contours on the other three images were obtained using \textit{SCSS-Net} predictions. The model was trained on three datasets with following annotations: (A) SPoCA, (B) Custom, (C) combination of SPoCA and Custom. The specific property of the model trained on a Custom dataset is that the model identifies larger areas of round shapes. On the other hand, the model trained on SPoCA dataset predicts too small regions. This was the reason why we decided to combine these datasets and create a third dataset from a combination of both.

\begin{figure}
    \centering
    \includegraphics[width=0.48\textwidth]{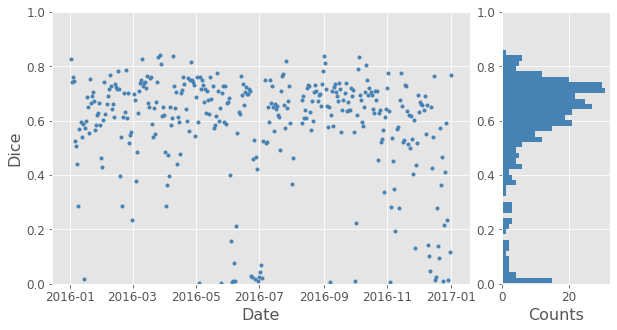}
    \caption{The values of Dice score for the model trained on SPoCA \& Custom dataset for each image in the Test set during the year 2016 (\textit{left}). Distribution of Dice score values in the Test set (\textit{right}). These panels extend the information of the averaged Dice score presented in Table \ref{tab:ar2}, last row.}
    \label{fig:ar_dice}
\end{figure}

As with the coronal holes, we tested our pre-trained models on segmentation masks obtained by the SPoCA reference annotations from the year 2016. The results of this experiment are presented in Table \ref{tab:ar2}. The visualization of \textit{SCSS-Net} performance over the whole test set is presented in Figure \ref{fig:ar_dice} and Figure \ref{fig:ar_area}. Examples of visual comparison of the same images' predictions is displayed in the appendix in Figures \ref{fig:ar_test_custom}-\ref{fig:ar_test_comb}. In general, the performance of the \textit{SCSS-Net} is satisfactory but not so good as for the coronal holes. The reason is imperfection of annotations needed for re-training of the model. Even though, it is presented that \textit{SCSS-Net} created for one type of coronal structures (coronal holes) can be easily re-trained for segmentation of another type of coronal structures (active regions). 

\begin{table}
	\centering
	\caption{Performance of \textit{SCSS-Net} pre-trained models for active region segmentation against the SPoCA reference annotations. The test set consists of input images from the year 2016, only.}
	\label{tab:ar2}
	\begin{tabular}{lccccr}
		\toprule
		\textbf{Train dataset} & \textbf{Test set} & \textbf{Dice} & \textbf{IoU} \\
		\midrule
		Custom  & 360 & 0.49 & 0.30  \\
		SPoCA & 360 & 0.68 & 0.51  \\
		SPoCA + Custom  & 360 & 0.64 & 0.47 \\
		\bottomrule
	\end{tabular}
\end{table}

\begin{figure}
    \centering
    \includegraphics[width=0.48\textwidth]{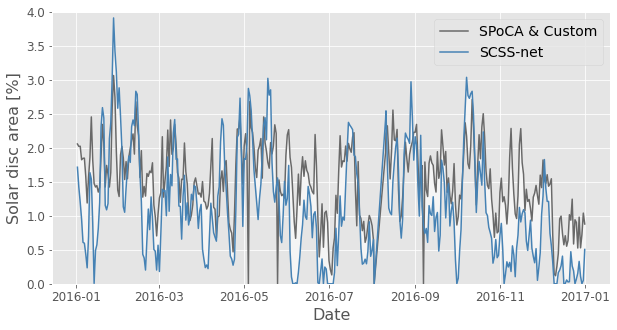}
    \caption{The area of segmented coronal holes in \% of total solar disc area by SPoCA \& Custom reference annotations and by \textit{SCSS-Net} model. The visualization is complementary to Figure \ref{fig:ar_dice}.}
    \label{fig:ar_area}
\end{figure}

\subsection{Comparison to other deep learning approaches}

Now we would like to discuss the comparison of our approach with the previously mentioned CNN-based methods in \cite{Illarionov2018} and \cite{jarolim2021multi}. From the general perspective, we aimed to design \textit{SCSS-Net} more universally and to investigate if it is able to adapt to different segmentation goals and to segment different types of coronal structures. In our study, we have investigated \textit{SCSS-Net} performance for coronal holes and active regions. Other papers focused on coronal holes only. In more detail, the \textit{SCSS-Net} architecture and the architecture of the model in \cite{Illarionov2018} are based on the same idea of a U-Net segmentation network \citep{ronneberger2015u}. However, in our case, we have a more robust version through more filters and regularizations. The main reason to use more filters is to provide a higher learning capacity. More regularization elements help adapt a larger learning space to a specific subdomain. Therefore, combining both ideas within \textit{SCSS-Net} adopts the same architecture (easily with simple retraining on respective annotations) to segment coronal holes and active regions. Another difference that can help to achieve a more robust result is that we applied different annotation sets. Thanks to their combination, final models can be more flexible and better trained to avoid biases within particular annotation sets.

The CHRONOS method, described recently in \cite{jarolim2021multi},  is inspired by the generative approaches and adopts an idea of a progressively growing network. In this case, this idea is applied to convolutional layers combined in similar encoder-decoder architecture as U-Net segmentation networks. The main difference is that it starts with smaller resolutions (8x8) and progressively increases resolution through the layers. One of the advantages of such architectures is that there is a smaller number of weights at the first layers of architecture, but the results based on IoU metric did not show more significant improvement to U-Net architectures. The authors used masks obtained from the SPoCA-CH module (updated SPoCA annotations, see \cite{Delouille2018}) for the training, which was then manually reviewed to remove the remaining filaments and invalid extractions. For additional testing, they used CATCH algorithm (Collection of Analysis Tools for Coronal Holes, \cite{catch}) for the identification of coronal holes. But only coronal holes between longitudes of [-400", 400"] in helioprojective coordinates were considered. They also used multiple SDO/AIA channels in order to provide more precise segmentation maps. 

For the comparison with our approach, we can discuss two aspects. First, we only used one 193\,\AA\, channel and the results were tested on different annotation sets. With default SPoCA annotations, we were able to achieve IoU 0.57 (see Table \ref{tab:ch2}). In their case, the same channel trained with manually reviewed SPoCA annotations achieved IoU 0.61. The main reason is only due to enhanced annotations, but we expected that our model learned to avoid problems identified by manual revision. The authors in the afore-mentioned paper used many spectral channels and identified the most important channels for segmentation regarding the multiple channel approach. We can see that globally best IoU is 0.63, only 2\% more than if 193\,\AA\, channel is used. Moreover, we used more annotation sets with different annotation biases and achieved consistently good segmentation results. Therefore, our use of the spectral channel seems efficient, and the architecture is robust for the segmentation of coronal holes. In the future, our architecture can be easily adapted for a multi-channel approach using a combination of several networks, with the selection of channels identified by the paper as most important for segmentation (based on metrics like IoU, recall, or precision).

\section{Conclusions}
\label{sec:conclusions}

The recent advances in computer vision might significantly automate and improve scientific analysis, mainly in fields with enormous data volumes \citep{Hausen2020}. Thanks to the fleet of space observatories, Heliophysics is a such field, and automation of data processing might be very beneficial for studies of Sun--Earth connections. We adopted a deep learning approach and developed \textit{SCSS-Net} a convolutional neural network for segmentation of solar corona structures. \textit{SCSS-Net} needs an image of the Sun as the input and produces a segmentation mask of the target structure as the output. With the transfer learning approach, where a similar task can be solved in different domains, the application of the presented model can be very straightforward for various segmentation tasks in solar atmosphere images. In this work, we have presented an end-to-end process from the pre-processing of input data and preparation of sources of the target ground truth annotations, through own design and optimization of the convolutional neural network and its training, to the evaluation of model performance and its testing on previously completely unseen data. We went through this process for corona holes and then re-trained the developed model for active regions. For both types of solar corona structures, quantitatively comparable results with generally used methods have been obtained. The main limitation of the model performance was the precision of the reference annotations. It means that even if the \textit{SCSS-Net} performance of segmentation precision was better than the segmentation precision by other methods it is difficult to evaluate the performance with confidence. Therefore, we have not only listed the evaluation metrics in the tables but also provide visual comparisons using several examples. In general, it will be very valuable if a master test dataset is created with correct and precise ground truth reference annotations that will be used as the main reference for benchmark of different approaches for segmentation of coronal structures. Indeed, this effort has already started as it is presented by \cite{Reiss2021}. Qualitatively, \textit{SCSS-Net} provides comparable results of segmentation of coronal holes as methods used in this study. Thanks to the discussed advantages of \textit{SCSS-Net}, it might be considered and employed in further review studies.

We plan to deploy the developed \textit{SCSS-Net} for at least the following two applications. Firstly, modeling of airglow intensity variation in the Earth's upper atmosphere as it is directly influenced by the amount of solar radiation in ultraviolet spectral range. The segmented solar corona structures will be additional input features for the model developed by \cite{Mackovjak2021}. The automatically obtained parameters like total area size, heliographic coordinates, and lifetime of specific coronal structures might provide very valuable supplemental information to the actually employed F10.7 index that represents general solar activity. For this application, the relative values of obtained parameters in time will be sufficient. The uncertainties possibly caused by imprecise \textit{SCSS-Net} inputs will not be significant. Secondly, automated search for specific structures in the solar corona. The events like slipping motions of flare loops \citep{Dudik2016} or saddle-shaped arcades of flare loops during eruptive flares \citep{Lorincik2021b} were investigated for several cases. All SDO/AIA images may be reviewed by the trained \textit{SCSS-Net} and many more occurrences of interesting structures might be possibly found for further studies.  

\section*{Acknowledgements}
Heliophysics studies presented in this work are supported by the Slovak VEGA grant agency project 2/0155/18 and also by the government of Slovakia through the ESA contracts No. 4000125330/18/NL/SC and No. 4000125987/18/NL/SC under the PECS (Plan for European Cooperating States). ESA Disclaimer: The view expressed herein can in no way be taken to reflect the official opinion of the European Space Agency. 

The studies of the deep learning approach presented in this work are supported by Slovak APVV research grant under contract No. APVV-16-0213 and Slovak VEGA research grant No. 1/0685/21. 

We are thankful to students of the Technical University of Ko\v{s}ice who helped us with the preparation of Custom annotations using the Zooniverse.org platform. This publication uses data generated via the Zooniverse.org platform, development of which is funded by generous support, including a Global Impact Award from Google, and by a grant from the Alfred P. Sloan Foundation. The authors are also thankful to the referee and editor for the remarks that led to improvements of the manuscript.

\section*{Data Availability}
AIA data are courtesy of NASA / SDO and the AIA science team. The data underlying results in this article will be shared on reasonable request to the corresponding author. The main results can be reproduced by Jupyter notebooks publicly available online\footnote{\url{https://github.com/space-lab-sk/scss-net}}.


\bibliographystyle{mnras}
\bibliography{scss-net_bib} 

\appendix
\section{Visualization of CH segmentation} 

\begin{figure}
    \centering
       \includegraphics[width=0.43\textwidth]{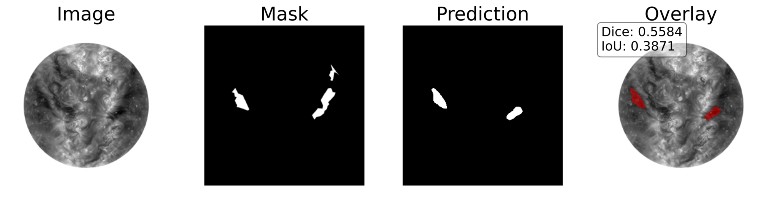}
        \includegraphics[width=0.43\textwidth]{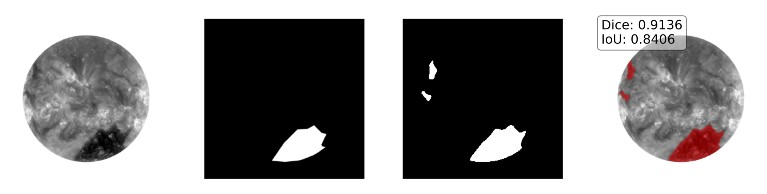}
    \caption{Segmentation of coronal holes by \textit{SCSS-Net} trained on dataset  with \textbf{Custom} annotations. The examples of two image sequences (top and bottom row) are selected from the validation dataset (not used in training process). Each image sequence consists of the input image for \textit{SCSS-Net}, ground truth (i.e. annotation mask obtained from Custom annotation), the output of \textit{SCSS-Net} (i.e. predicted segmentation mask), and the predicted segmentation mask in red as an overlay on the input image. The values of Dice and IoU metrics are displayed for each particular prediction.}
    \label{fig:val_custom}
    
        \includegraphics[width=0.43\textwidth]{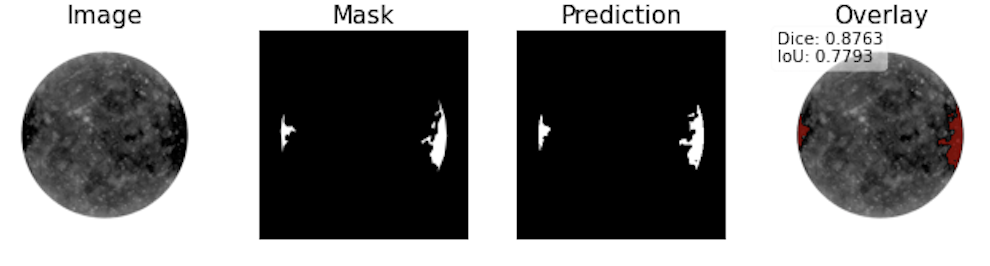}
        \includegraphics[width=0.43\textwidth]{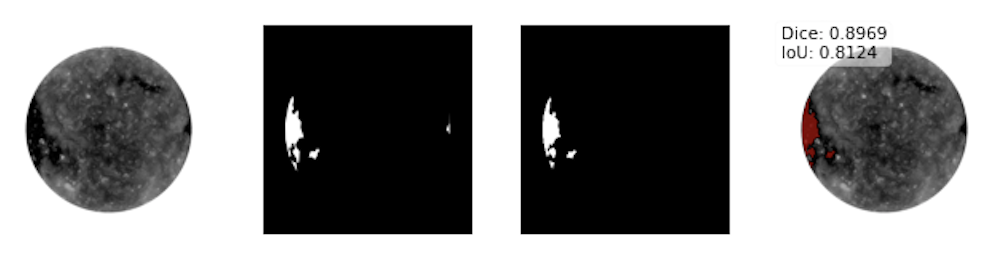}
    \caption{Segmentation of coronal holes by \textit{SCSS-Net} trained on the dataset with \textbf{SPoCA} annotations. For more details refer to the caption of Figure \ref{fig:val_custom}.}
    \label{fig:val_spoca}
    
        \includegraphics[width=0.43\textwidth]{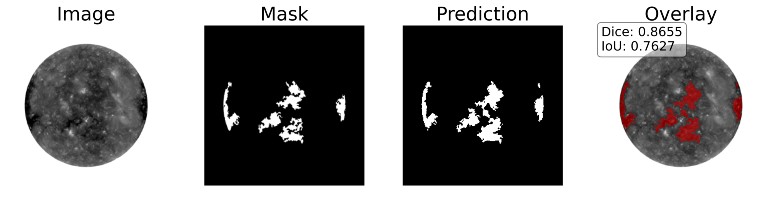}
        \includegraphics[width=0.43\textwidth]{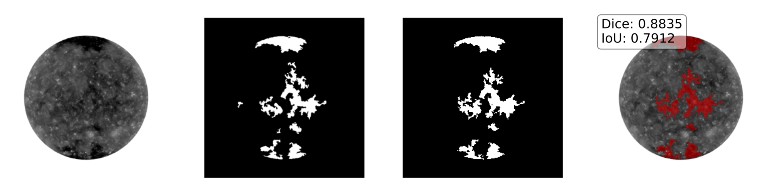}
    \caption{Segmentation of coronal holes by \textit{SCSS-Net} trained on the dataset with \textbf{CHIMERA} annotations. For more details refer to the caption of Figure \ref{fig:val_custom}.}
    \label{fig:val_chimera} 
    
        \includegraphics[width=0.43\textwidth]{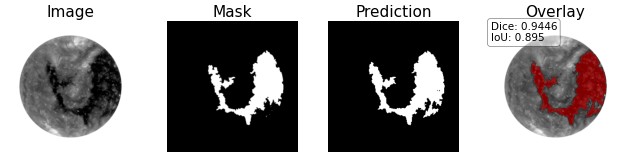}
        \includegraphics[width=0.43\textwidth]{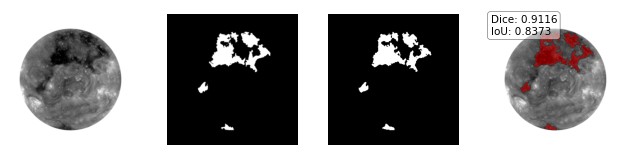}
    \caption{Segmentation of coronal holes by \textit{SCSS-Net} trained on the dataset with \textbf{Region Growth} annotations. For more details refer to the caption of Figure \ref{fig:val_custom}.}
    \label{fig:val_region}
\end{figure}

\begin{figure}
    \centering
        \includegraphics[width=0.43\textwidth]{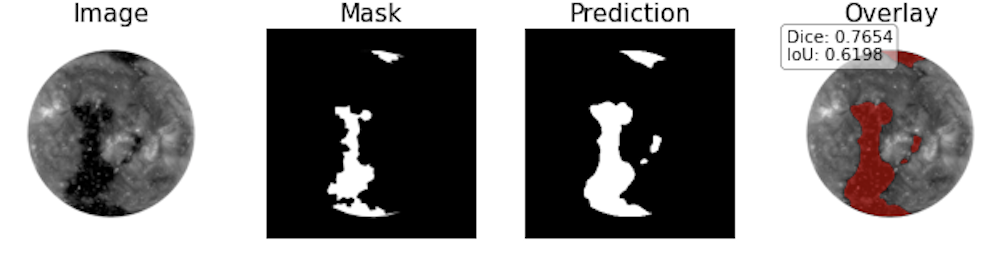}
        \includegraphics[width=0.43\textwidth]{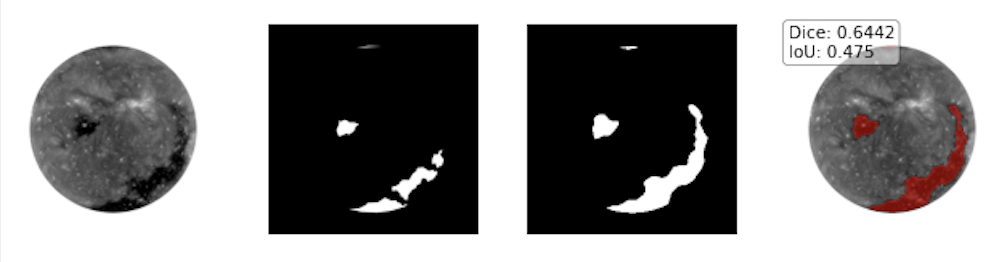}  
    \caption{Results of the \textbf{Custom} pre-trained model (coronal holes) with reference masks obtained using the SPoCA algorithm. The examples of two image sequences (top and bottom row) are selected from the test dataset from year 2017. Each image sequence consists of the input image for \textit{SCSS-Net}, reference mask, the output of \textit{SCSS-Net} (i.e. predicted segmentation mask), and the predicted segmentation mask in red as an overlay on the input image. The values of Dice and IoU metrics are displayed for each particular prediction.}
    \label{fig:test1_custom}
    
        \includegraphics[width=0.43\textwidth]{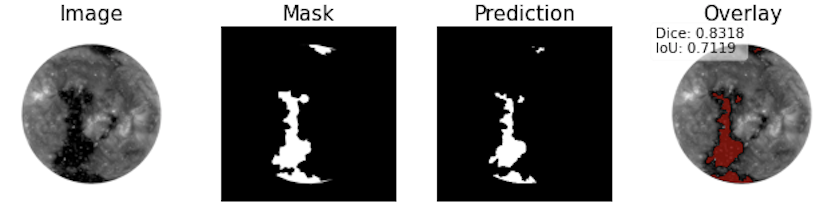}
        \includegraphics[width=0.43\textwidth]{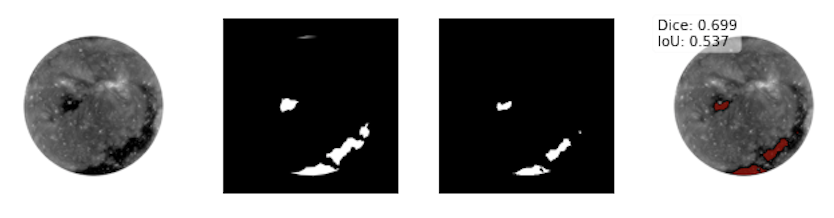}  
    \caption{Results of the \textbf{SPoCA} pre-trained model (coronal holes) with reference masks obtained using the SPoCA algorithm on AIA images from 2017. For more details refer to the caption of Figure \ref{fig:test1_custom}.}
     \label{fig:test1_spoca}
     
        \includegraphics[width=0.43\textwidth]{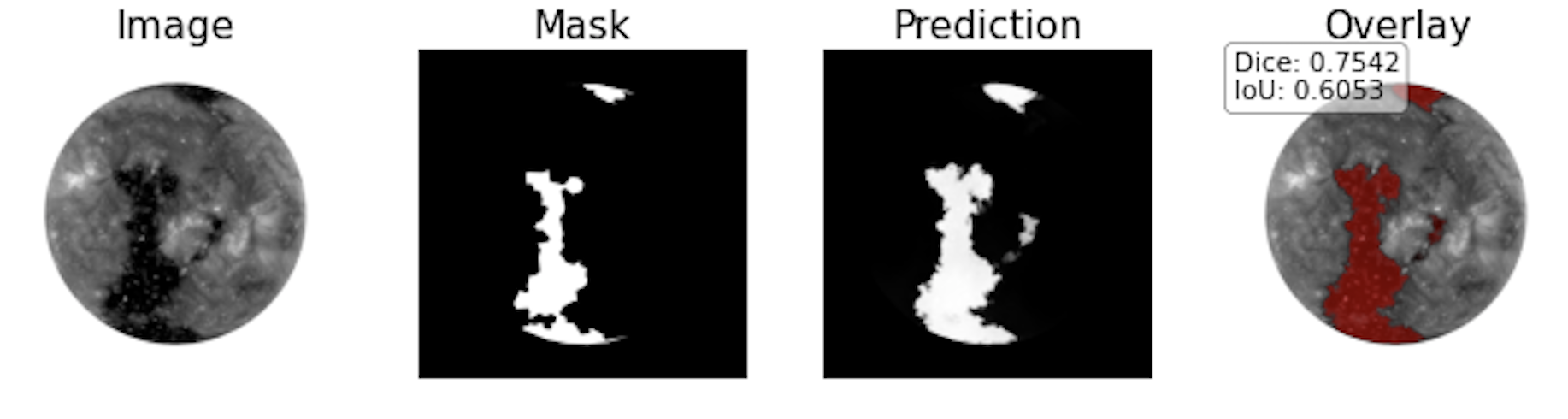}
        \includegraphics[width=0.43\textwidth]{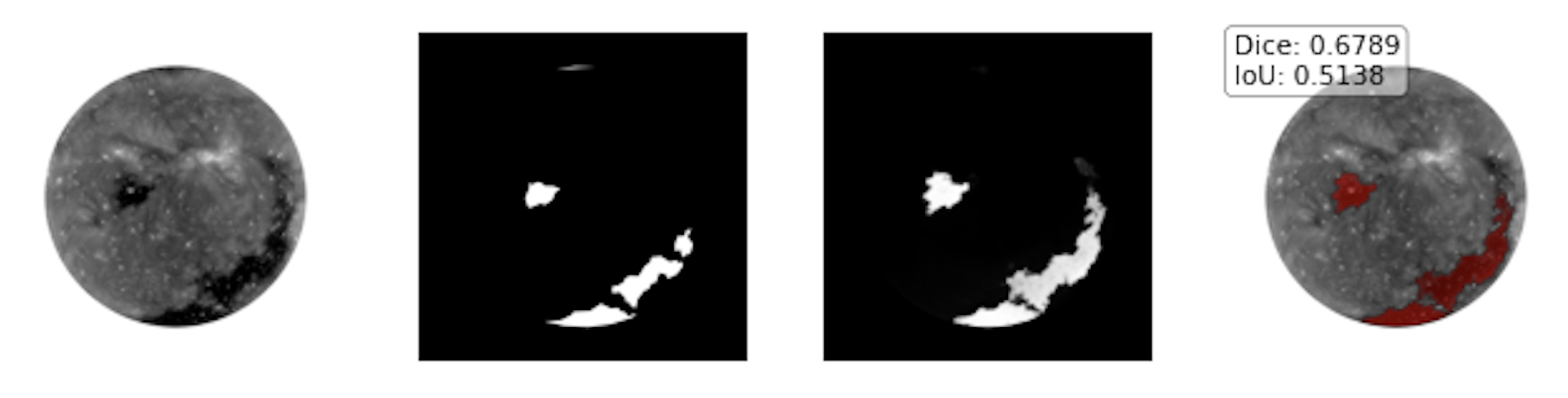}  
    \caption{Results of the \textbf{CHIMERA} pre-trained model (coronal holes) with reference masks obtained using the SPoCA algorithm on AIA images from 2017. For more details refer to the caption of Figure \ref{fig:test1_custom}.}
     \label{fig:test1_chimera}
     
        \includegraphics[width=0.43\textwidth]{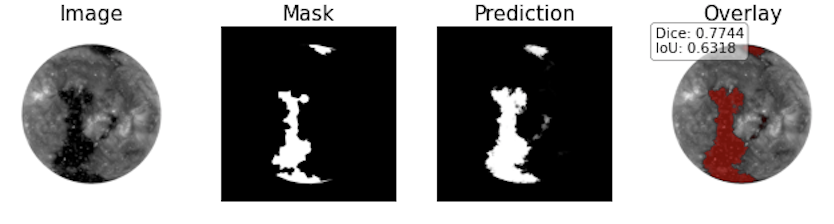}
        \includegraphics[width=0.43\textwidth]{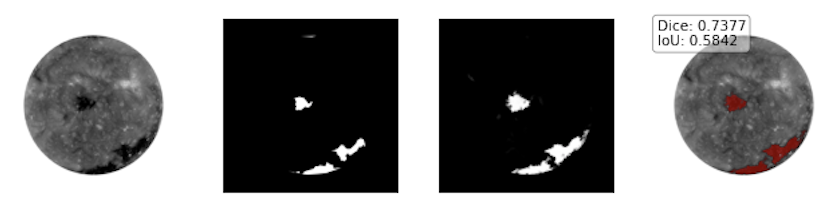}  
    \caption{Results of the \textbf{Region Growth} pre-trained model (coronal holes) with reference masks obtained using the SPoCA algorithm on AIA images from 2017. For more details refer to the caption of Figure \ref{fig:test1_custom}.}
     \label{fig:test1_region}
\end{figure}

 \begin{figure}
    \centering
        \includegraphics[width=0.43\textwidth]{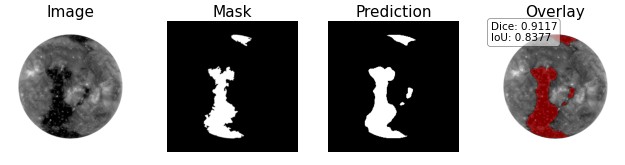}
        \includegraphics[width=0.43\textwidth]{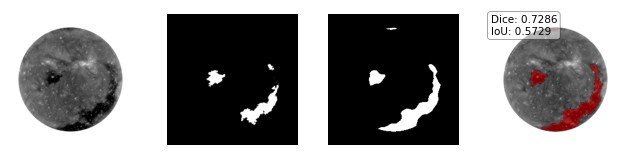}  
    \caption{Results of the \textbf{Custom} pre-trained model (coronal holes) with reference masks obtained using the  the Region Growth algorithm. The examples of two image sequences (top and bottom row) are selected from the test dataset from year 2017. Each image sequence consists of the input image for \textit{SCSS-Net}, reference mask, the output of \textit{SCSS-Net} (i.e. predicted segmentation mask), and the predicted segmentation mask in red as an overlay on the input image. The values of Dice and IoU metrics are displayed for each particular prediction.}
    \label{fig:test2_custom}
    
        \includegraphics[width=0.43\textwidth]{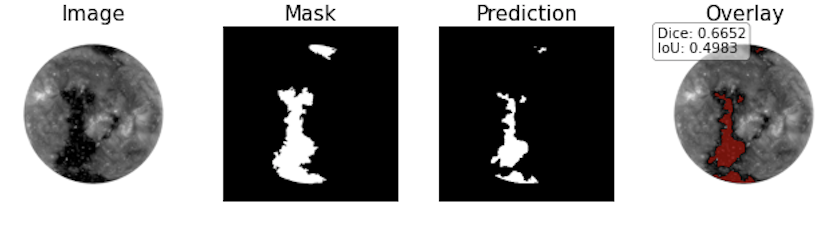}
        \includegraphics[width=0.43\textwidth]{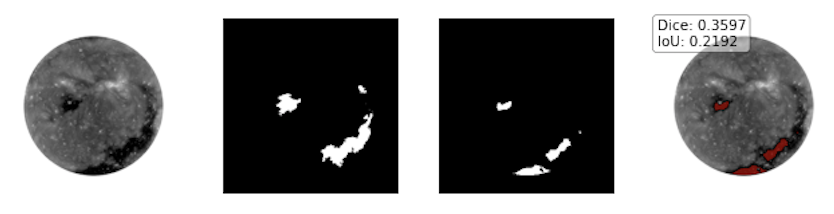}  
    \caption{Results of the \textbf{SPoCA} pre-trained model (coronal holes) with reference masks obtained using the Region Growth algorithm on AIA images from 2017. For more details refer to the caption of Figure \ref{fig:test2_custom}.}
    \label{fig:test2_spoca}
    
        \includegraphics[width=0.43\textwidth]{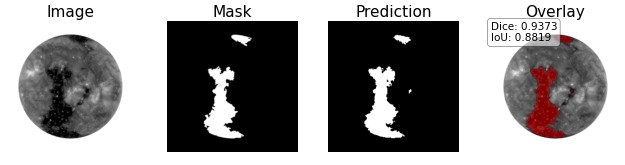}
        \includegraphics[width=0.43\textwidth]{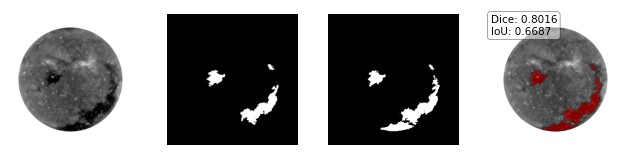}  
    \caption{Results of the \textbf{CHIMERA} pre-trained model (coronal holes) with reference masks obtained using the Region Growth algorithm on AIA images from 2017. For more details refer to the caption of Figure \ref{fig:test2_custom}.}
    \label{fig:test2_chimera}
    
        \includegraphics[width=0.43\textwidth]{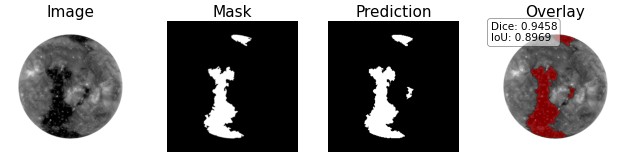}
        \includegraphics[width=0.43\textwidth]{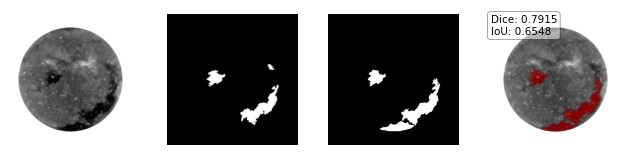}  
    \caption{Results of the \textbf{Region Growth} pre-trained model (coronal holes) with reference masks obtained using the Region Growth algorithm on AIA images from 2017. For more details refer to the caption of Figure \ref{fig:test2_custom}.}
    \label{fig:test2_region}
\end{figure}

\section{Visualization of AR segmentation}

\begin{figure}
    \centering
        \includegraphics[width=0.43\textwidth]{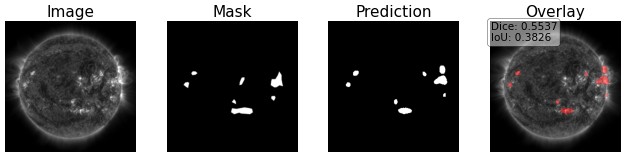}
        \includegraphics[width=0.43\textwidth]{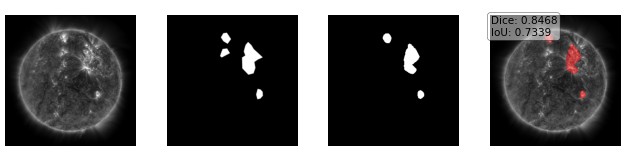}
    \caption{Segmentation of active regions by \textit{SCSS-Net} trained on the dataset  with \textbf{Custom} annotations. The examples of two image sequences (top and bottom row) are selected from the validation dataset. Each image sequence consists of input image for \textit{SCSS-Net}, ground truth (i.e. annotation mask obtained from Custom annotation), the output of \textit{SCSS-Net} (i.e. predicted segmentation mask), and predicted segmentation mask in red as an overlay on the input image. The values of Dice and IoU metrics are displayed for each particular prediction.}
    \label{fig:ar_val_custom}
    
        \includegraphics[width=0.43\textwidth]{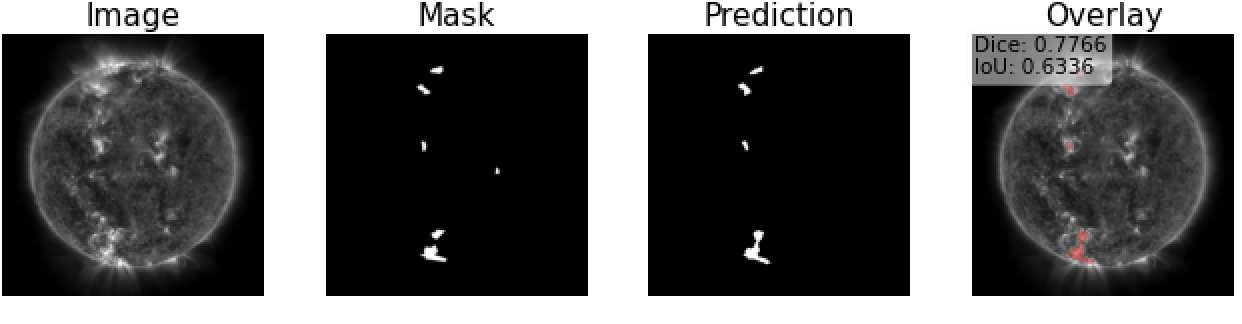}
        \includegraphics[width=0.43\textwidth]{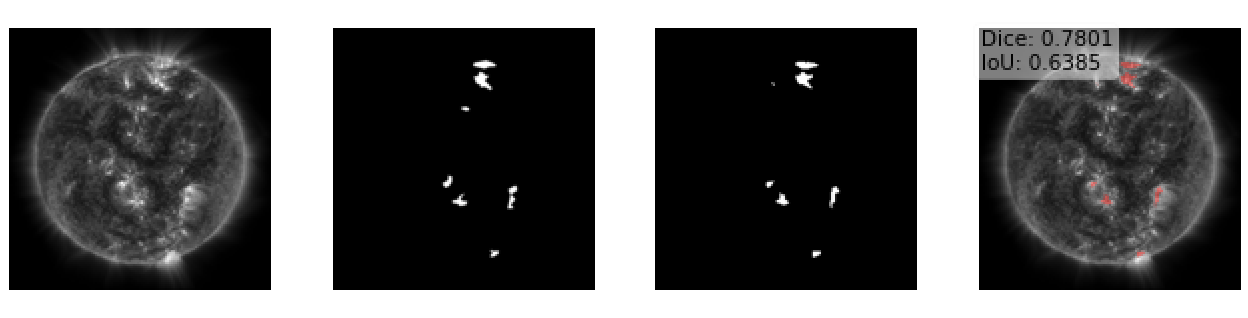}
    \caption{Segmentation of coronal holes by \textit{SCSS-Net} trained on the dataset with \textbf{SPoCA} annotations. For more details refer to the caption of Figure \ref{fig:ar_val_custom}.}
    \label{fig:ar_val_spoca}   
 
        \includegraphics[width=0.43\textwidth]{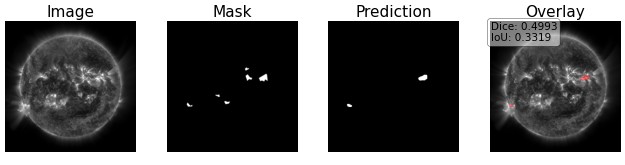}
        \includegraphics[width=0.43\textwidth]{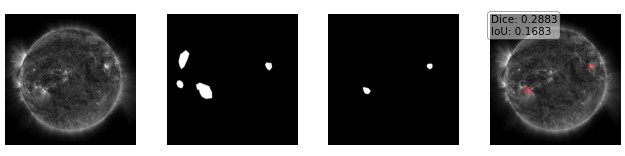}
    \caption{Segmentation of coronal holes by \textit{SCSS-Net} trained on dataset with \textbf{combination of SPoCA and Custom} annotations. For more details refer to the caption of Figure \ref{fig:ar_val_custom}.}
    \label{fig:ar_val_comb}
\end{figure}

\begin{figure}
    \centering
        \includegraphics[width=0.43\textwidth]{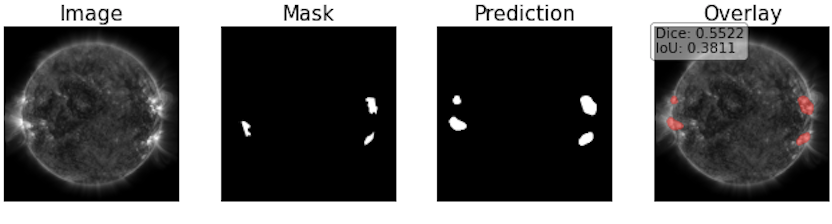}
        \includegraphics[width=0.43\textwidth]{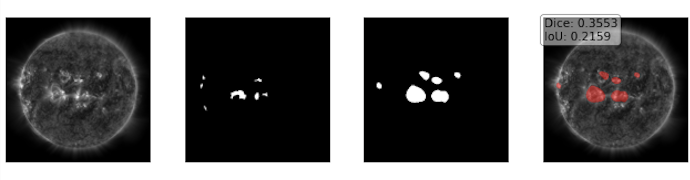}  
    \caption{Results of the \textbf{Custom} pre-trained model (active regions) on segmentation masks obtained using the SPoCA algorithm on AIA images from 2016.}
    \label{fig:ar_test_custom}

       \includegraphics[width=0.43\textwidth]{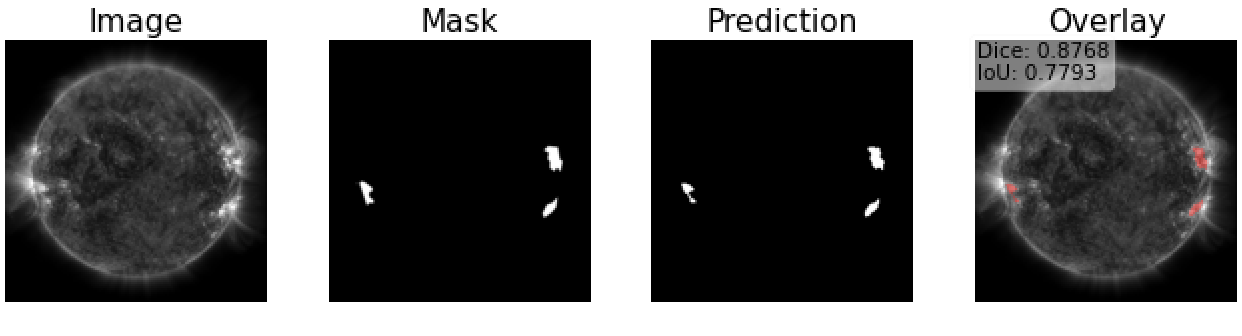}
        \includegraphics[width=0.43\textwidth]{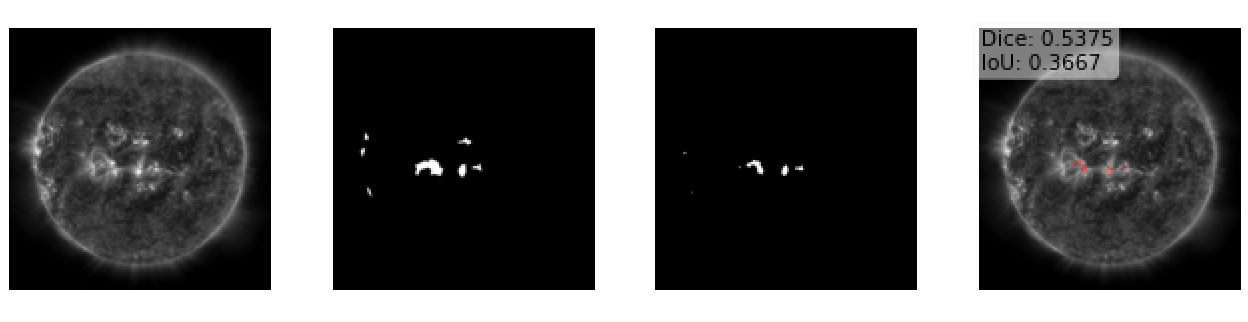}  
    \caption{Results of the \textbf{SPoCA} pre-trained model (active regions) on segmentation masks obtained using the SPoCA algorithm on AIA images from 2016.}
    \label{fig:ar_test_spoca}    
 
        \includegraphics[width=0.43\textwidth]{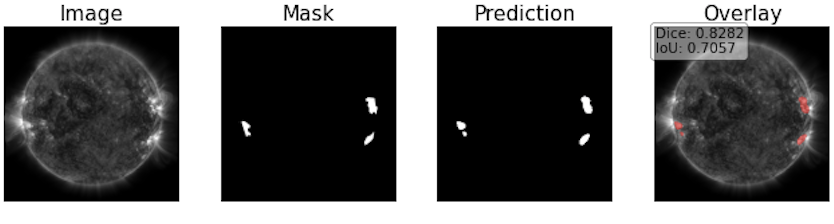}
        \includegraphics[width=0.43\textwidth]{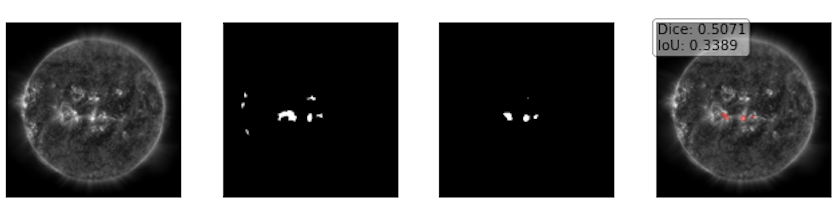}  
    \caption{Results of the \textbf{combination of SPoCA and Custom} pre-trained model (active regions) on segmentation masks obtained using the SPoCA algorithm on AIA images from 2016.}  
    \label{fig:ar_test_comb}   
\end{figure}


\bsp	
\label{lastpage}
\end{document}